\documentclass[twocolumn]{aastex631}

\usepackage{amsmath}
\usepackage{apjfonts, natbib}

\received{April 26, 2024}
\revised{July 10, 2024}
\accepted{July 15, 2024}

\begin{document}

\title{The unluckiest star: A spectroscopically confirmed repeated partial tidal disruption event AT\,2022dbl}

\author[0000-0003-4959-1625]{Zheyu Lin}
\affiliation{Department of Astronomy, University of Science and Technology, Hefei, 230026, China; linzheyu@mail.ustc.edu.cn, jnac@ustc.edu.cn, twang@ustc.edu.cn, xkong@ustc.edu.cn}
\affiliation{School of Astronomy and Space Sciences,
University of Science and Technology of China, Hefei, 230026, China}

\author[0000-0002-7152-3621]{Ning Jiang}
\affiliation{Department of Astronomy, University of Science and Technology, Hefei, 230026, China; linzheyu@mail.ustc.edu.cn, jnac@ustc.edu.cn, twang@ustc.edu.cn, xkong@ustc.edu.cn}
\affiliation{School of Astronomy and Space Sciences,
University of Science and Technology of China, Hefei, 230026, China}

\author[0000-0002-1517-6792]{Tinggui Wang}
\affiliation{Department of Astronomy, University of Science and Technology, Hefei, 230026, China; linzheyu@mail.ustc.edu.cn, jnac@ustc.edu.cn, twang@ustc.edu.cn, xkong@ustc.edu.cn}
\affiliation{School of Astronomy and Space Sciences,
University of Science and Technology of China, Hefei, 230026, China}
\affiliation{Institute of Deep Space Sciences, Deep Space Exploration Laboratory, Hefei 230026, China}

\author[0000-0002-7660-2273]{Xu Kong}
\affiliation{Department of Astronomy, University of Science and Technology, Hefei, 230026, China; linzheyu@mail.ustc.edu.cn, jnac@ustc.edu.cn, twang@ustc.edu.cn, xkong@ustc.edu.cn}
\affiliation{School of Astronomy and Space Sciences,
University of Science and Technology of China, Hefei, 230026, China}
\affiliation{Institute of Deep Space Sciences, Deep Space Exploration Laboratory, Hefei 230026, China}

\author[0000-0002-4562-7179]{Dongyue Li}
\affiliation{National Astronomical Observatories, Chinese Academy of Sciences, Beijing, 100101, China}

\author[0009-0007-7292-8392]{Han He}
\affiliation{Department of Astronomy, School of Physics and Technology, Wuhan University, Wuhan 430072, China}


\author[0000-0003-4225-5442]{Yibo~Wang}
\affiliation{Department of Astronomy, University of Science and Technology, Hefei, 230026, China; linzheyu@mail.ustc.edu.cn, jnac@ustc.edu.cn, twang@ustc.edu.cn, xkong@ustc.edu.cn}
\affiliation{School of Astronomy and Space Sciences,
University of Science and Technology of China, Hefei, 230026, China}

\author[0000-0003-3824-9496]{Jiazheng Zhu}
\affiliation{Department of Astronomy, University of Science and Technology, Hefei, 230026, China; linzheyu@mail.ustc.edu.cn, jnac@ustc.edu.cn, twang@ustc.edu.cn, xkong@ustc.edu.cn}
\affiliation{School of Astronomy and Space Sciences,
University of Science and Technology of China, Hefei, 230026, China}

\author[0009-0006-2521-033X]{Wentao Li}
\affiliation{Department of Astronomy, University of Science and Technology, Hefei, 230026, China; linzheyu@mail.ustc.edu.cn, jnac@ustc.edu.cn, twang@ustc.edu.cn, xkong@ustc.edu.cn}
\affiliation{School of Astronomy and Space Sciences,
University of Science and Technology of China, Hefei, 230026, China}

\author[0000-0002-9092-0593]{Ji-an Jiang}
\affiliation{Department of Astronomy, University of Science and Technology, Hefei, 230026, China; linzheyu@mail.ustc.edu.cn, jnac@ustc.edu.cn, twang@ustc.edu.cn, xkong@ustc.edu.cn}
\affiliation{School of Astronomy and Space Sciences,
University of Science and Technology of China, Hefei, 230026, China}
\affiliation{National Astronomical Observatory of Japan, 2-21-1 Osawa, Mitaka, Tokyo 181-8588, Japan}

\author[0000-0003-2091-622X]{Avinash Singh}
\affiliation{Hiroshima Astrophysical Science Center, Hiroshima University, Higashi-Hiroshima, Hiroshima 739-8526, Japan}

\author[0000-0002-0525-0872]{Rishabh Singh Teja}
\affiliation{Indian Institute of Astrophysics, II Block, Koramangala, Bengaluru-560034, Karnataka, India}
\affiliation{Pondicherry University, R.V. Nagar, Kalapet, Pondicherry-605014, UT of Puducherry, India}

\author[0000-0002-6688-0800]{D. K. Sahu}
\affiliation{Indian Institute of Astrophysics, II Block, Koramangala, Bengaluru-560034, Karnataka, India}

\author[0000-0002-2006-1615]{Chichuan Jin}
\affiliation{National Astronomical Observatories, Chinese Academy of Sciences, Beijing, 100101, China}

\author[0000-0003-2611-7269]{Keiichi Maeda}
\affiliation{Department of Astronomy, Kyoto University, Kitashirakawa-Oiwake-cho, Sakyo-ku, Kyoto 606-8502, Japan}

\author[0000-0001-7689-6382]{Shifeng Huang}
\affiliation{Department of Astronomy, University of Science and Technology, Hefei, 230026, China; linzheyu@mail.ustc.edu.cn, jnac@ustc.edu.cn, twang@ustc.edu.cn, xkong@ustc.edu.cn}
\affiliation{School of Astronomy and Space Sciences,
University of Science and Technology of China, Hefei, 230026, China}



\begin{abstract}
The unluckiest star orbits a supermassive black hole elliptically. Every time it reaches the pericenter, it shallowly enters the tidal radius and gets partially tidal disrupted, producing a series of flares. Confirmation of a repeated partial tidal disruption event (pTDE) requires not only evidence to rule out other types of transients, but also proof that only one star is involved, as TDEs from multiple stars can also produce similar flares.  In this letter, we report the discovery of a repeated pTDE, AT\,2022dbl. In a quiescent galaxy at $z=0.0284$, two separate optical/UV flares have been observed in 2022 and 2024, with no bright X-ray, radio or mid-infrared counterparts. Compared to the first flare, the second flare has a similar blackbody temperature of $\sim$26,000 K, slightly lower peak luminosity, and slower rise and fall phases. Compared to the ZTF TDEs, their blackbody parameters and light curve shapes are all similar. The spectra taken during the second flare show a steeper continuum than the late-time spectra of the previous flare, consistent with a newly risen flare. More importantly, the possibility of two independent TDEs can be largely ruled out because the optical spectra taken around the peak of the two flares exhibit highly similar broad Balmer, N \textsc{iii} and possible He \textsc{ii} emission lines, especially the extreme $\sim$4100 \AA\ emission lines. This represents the first robust spectroscopic evidence for a repeated pTDE, which can soon be verified by observing the third flare, given its short orbital period.
\end{abstract}

\keywords{}


\section{Introduction} \label{sec:intro}
An unlucky star passes too close to a supermassive black hole (SMBH). It gets tidally torn apart and produces a luminous flare. In this case, a tidal disruption event (TDE) occurs \citep{Hills1975, Rees1988}. The discovery of these TDEs has been limited by the relatively low occurrence rate of about $10^{-4}-10^{-5}$ galaxy$^{-1}$ yr$^{-1}$ \citep[e.g.,][]{Wang2004,Stone2016,vanVelzen2020,Yao2023,Teboul2024}. From the late 1990s to the late 2000s, only $\sim$10 TDEs had been discovered. Most of them are bright in the X-ray bands \citep[e.g.,][]{Bade1996,Komossa1999,Esquej2007}, which are in concordance with the early theoretical prediction that the emission peaks in the extreme-UV to soft X-ray bands \citep[e.g.,][]{Rees1990,Cannizzo1990,Ulmer1999}. In the recent decade, however, more TDEs have been discovered by the wide-field optical surveys, such as the All-Sky Automated Survey for Supernovae (ASAS-SN), the Asteroid Terrestrial-impact Last Alert System (ATLAS) survey and the Zwicky Transient Facility (ZTF), and the current number of TDEs has greatly increased to $\sim$100 \citep[e.g.,][]{Gezari2021, Hammerstein2023, Yao2023}. Most of these TDEs are bright in optical/UV wavelengths but much fainter in X-ray, contrary to those earlier discovered TDEs. The origin of optical/UV emission is still under debate \citep[e.g.,][]{Loeb1997,Piran2015,Metzger2016,Dai2018,Lu2020,Liu2021,Thomsen2022}, awaiting the definitive observational evidence. As a result, the identification of optical/UV TDEs is empirical, relying on the features of the former samples.

A luckier star has a shallower encounter with an SMBH. Only part of it gets tidally disrupted and produces a similar flare. In this case, a partial tidal disruption event (pTDE) happens. The shallowness of the encounter is usually defined by the ratio of the tidal radius and the pericenter, or the penetration factor, $\beta\equiv R_{\rm t}/R_{\rm p}$. Numerical simulations have found that the critical $\beta$ values for the onset of the pTDE and the full TDE depend on the density profile of the star \citep[e.g.,][]{Guillochon2013,Law-Smith2017,Ryu2020II}. The event rate for pTDEs is predicted to be comparable or even higher than that of full TDEs \citep[e.g.,][]{Stone2016,Stone2020,Ryu2020IV,Chen2021,Zhong2022}, providing a boost to the total TDE rate. However, distinguishing pTDEs from full TDEs is difficult, as the luminosity is not only determined by $\beta$ or the disrupted mass, but also depends on other parameters such as the radiation efficiency, the BH mass and stellar properties.

Sometimes, this stroke of luck instead leads to tragedy. The unluckiest star initially has an elliptical orbit. Each time it approaches the pericenter, it experiences partial disruption, producing a series of flares. In this special case, a repeated pTDE occurs. The elliptical orbit of the star is possibly created by the Hills mechanism, in which a stellar binary passes by an SMBH and gets broken into a hyper-velocity star and a tightly bound star \citep{Hills1988,Cufari2022,Lu2023}. Repeated pTDEs can provide precious evidence for the existence of pTDEs. However, the confirmation of repeated pTDEs can be complicated by other possible scenarios, such as a double TDE caused by an extremely close encounter between a stellar binary and either an SMBH \citep{Mandel2015} or a milliparsec-scale SMBH binary \citep{Wu2018}. Alternatively, multiple independent TDEs could be supported by an enhanced TDE rate, due to the concentrated nuclear stellar profile, e.g., in post-starburst galaxies \citep[e.g.,][]{Arcavi2014,Hammerstein2021,Bortolas2022,Wang2024} or galaxies with nuclear star clusters \citep{Pfister2020}. Despite the challenges, several candidates for repeated pTDE have been reported, e.g., IC 3599 \citep{Campana2015}, ASASSN-14ko \citep{Payne2021,Payne2022,Payne2023,Huang2023}, Swift J023017.0+283603 \citep{Evans2023,Guolo2024}, eRASSt J045650.3–203750 \citep{Liu2023,Liu2024}, AT\,2018fyk \citep{Wevers2023,Pasham2024}, RX J133157.6–324319.7 \citep{Hampel2022,Malyali2023} and AT\,2020vdq \citep{Somalwar2023}. The great diversity of the flaring intervals, bands and shapes (listed in Table \ref{tab:pTDE}) among these sources calls for additional theoretical efforts.


In this letter, we report the discovery of a new recurring flare at the position of AT\,2022dbl (also known as AT\,2018mac, ZTF18aabdajx, ASASSN-22ci). It follows the dissipation of the tidal disruption flare that rose two years ago. Photometric and spectroscopic follow-up observations have been conducted since this discovery, confirming that this flare is also the result of a TDE. Its extreme $\sim$4100 \AA\ emission line resembles the last flare, providing vital evidence for a repeated pTDE. 

Since the discovery of the recurring flare on January 22, 2024, we have performed extensive photometric and spectroscopic observations. Meanwhile, we have also collected historical photometric and spectroscopic data to provide a comprehensive view of this event.

The letter is organized as follows. In the appendix (Section \ref{sec:obs}), we present the observations and data reduction procedures. In Section \ref{sec:ana}, we analyze the host galaxy, the historical and recent photometric evolution in UV, optical and X-ray bands, as well as the optical spectra. In Section \ref{sec:discuss}, we discuss the possible origins of AT\,2022dbl and compare it with other repeated pTDEs. A final summary is given in Section \ref{sec:conclude}. All errors marked with ``$\pm$'' represent the 1-$\sigma$ confidence intervals. We assume a flat cosmology with $H_{0} =70$ km~s$^{-1}$~Mpc$^{-1}$ and $\Omega_{\Lambda} = 0.7$. For the extinction correction, we use the extinction law of \citet{Fitzpatrick1999}, the standard extinction curve with $R_V=A_V/ E(B-V)=3.1$ \citep{Osterbrock2006} and adopt a Galactic extinction of $E(B-V) = 0.0159$ mag \citep{Planck2016}. All magnitudes are in the AB system \citep{Oke1974}.

\section{Data Analysis} \label{sec:ana}

\subsection{Host Galaxy} \label{subsec:hostana}
\begin{figure*}[htb!]
    \centering
    \epsscale{0.8}
    \plotone{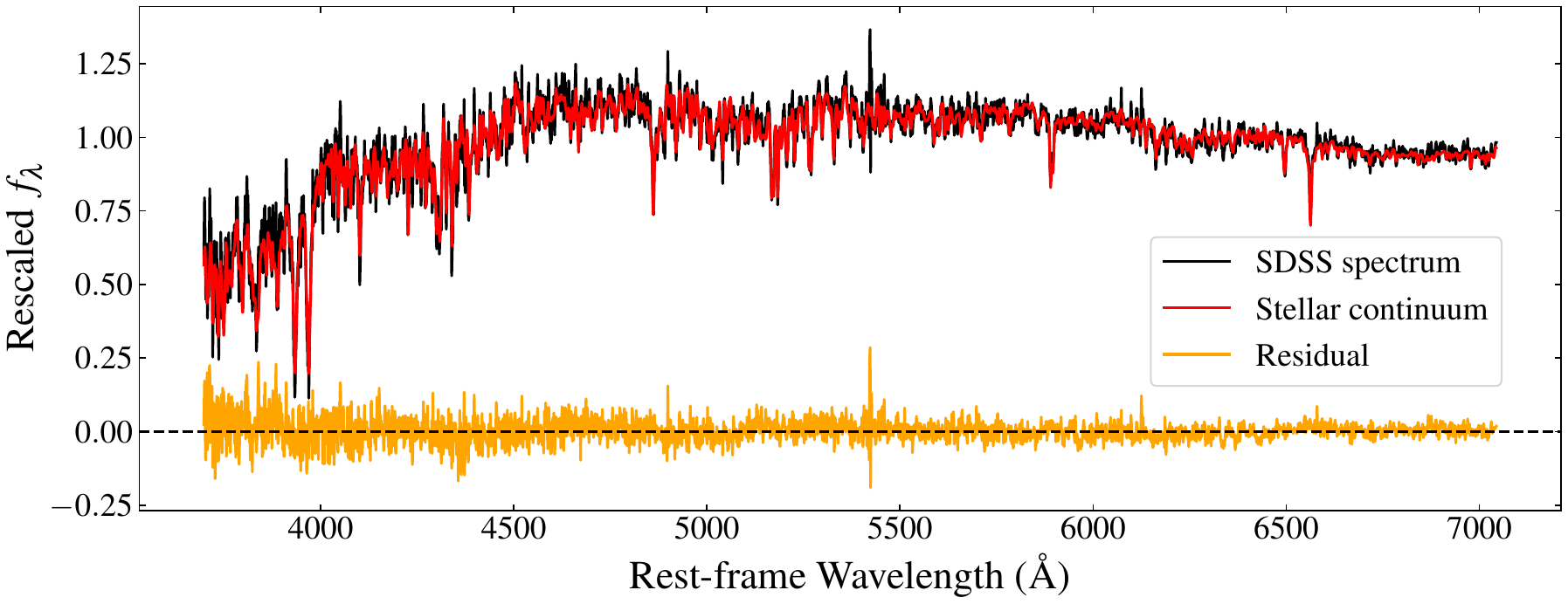}
        \caption{The \texttt{pPXF} fitting result of the SDSS host-galaxy spectrum. The residual shows no clear emission line, implying that the host galaxy should not be an AGN.}
    \label{fig:ppxf}
\end{figure*}

The host galaxy SDSS J122045.04+493304.6 has an early SDSS spectrum. The redshift is $z=0.02840\pm0.00001$. To examine the possible AGN activity, we fit the SDSS spectrum by the penalized pixel-fitting (\texttt{pPXF}) software \citep{Cappellari2023}. We adopt the flexible stellar population synthesis (FSPS) model templates \citep{Conroy2009}, and mask the common galaxy emission and absorption lines before fitting the stellar continuum. The residual is obtained after subtracting the best-fit stellar continuum. As shown in Figure \ref{fig:ppxf}, the residual shows no clear emission line, which means that the host-galaxy spectrum can be fitted by a single stellar component. Therefore, the host galaxy should not be an AGN.

The velocity dispersion derived from the stellar continuum is $\sigma=66.92\pm2.71$ km s$^{-1}$. Using the relation of \citet{Kormendy2013}, we derive a black hole mass of log $(M_{\rm BH}/M_\odot)=6.40\pm0.33$.


The spectrum displays Balmer absorption line series of H$\alpha$, H$\beta$, H$\gamma$ and H$\delta$. We derive a H$\alpha$ equivalent width (EW) emission of 0.015 $\pm$ 0.020 \AA\ and a Lick H$\delta_{\rm A}$ index of 2.09 $\pm$ 0.44 \AA. These parameters agree with the criteria of the quiescent Balmer-strong (QBS) galaxy: H$\alpha$ EW emission $<3$ \AA\ and H$\delta_{\rm A}>1.31$ \AA\ \citep{French2016}.

\subsection{UV/Optical Photometric Analysis} \label{subsec:uvoptana}

\begin{figure*}[htb!]
    \centering
    \epsscale{1.}
    \plotone{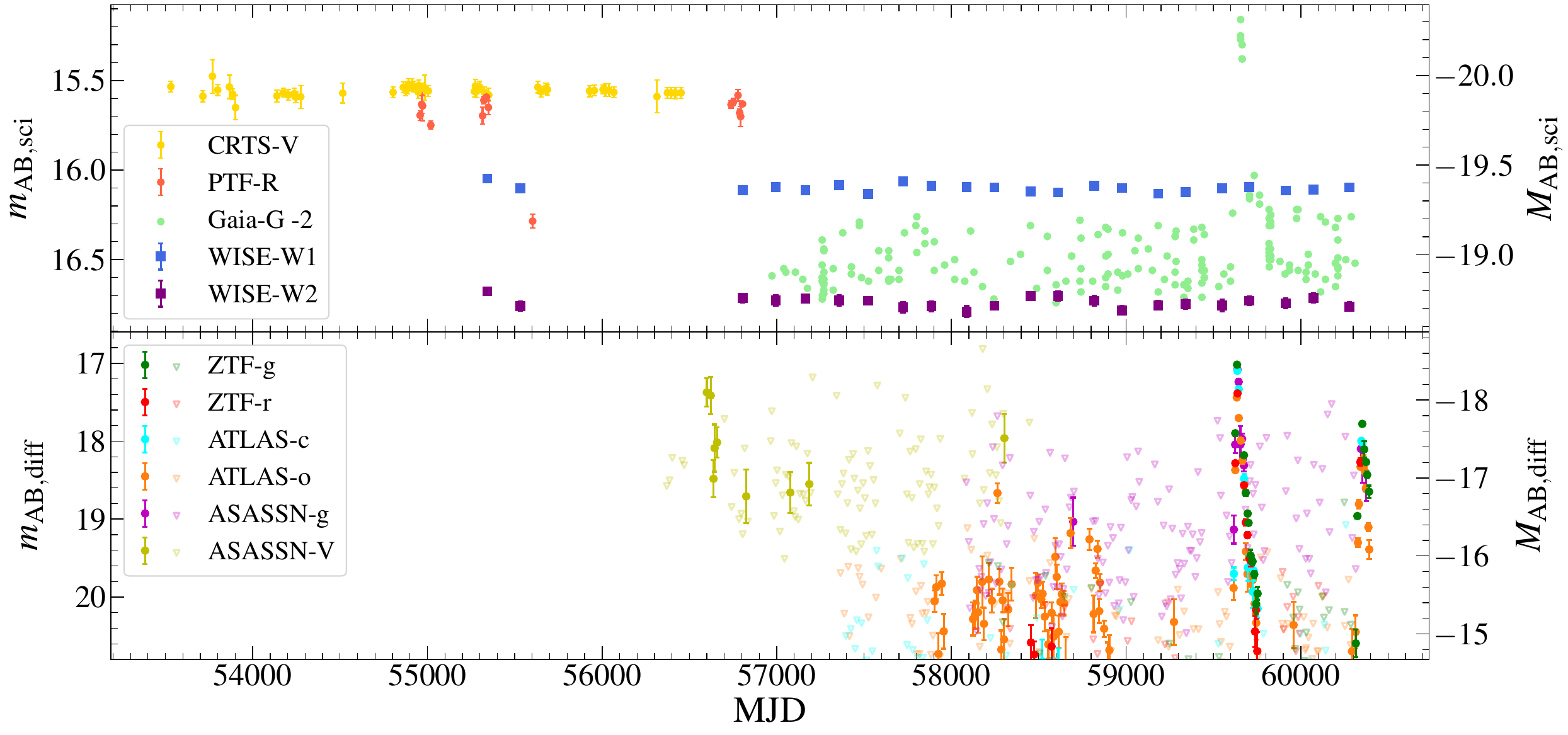}
    \caption{The historical light curves of the position of AT\,2022dbl. \textbf{Top panel:} The non-host-subtracted light curves of three optical surveys: CRTS (V band), PTF (R band) and Gaia (G band), along with the mid-infrared (MIR) WISE survey (W1 and W2 bands). \textbf{Bottom panel:} The host-subtracted light curves of three optical surveys: ZTF ($g$ and $r$ bands), ATLAS ($c$ and $o$ bands) and ASAS-SN ($g$ and V bands). To improve the SNR, we binned the data into 10-day bins for all optical bands except for Gaia-G, and into approximately half-year bins for W1 and W2 bands. Note that a potential early flare is displayed in the ASAS-SN V-band light curve, we discuss its reliability in Section \ref{subsec:preflare}.}
    \label{fig:preflare}
\end{figure*}

\subsubsection{Historical Variability}\label{subsec:preflare}
To check if there was any variability before the 2022 outburst, we first query the differential photometric data of ZTF ($g$ and $r$ bands), ATLAS ($c$ and $o$ bands) and ASAS-SN ($g$ and V bands), as mentioned in Section \ref{subsec:ztf} and \ref{subsec:atlasasas}. In addition, we query the Gaia Photometric Science Alerts (G band) and archival CRTS (V band) and PTF (R band) catalogs, as introduced in Section \ref{subsec:arch}. In addition, we query the AllWISE and NEOWISE catalogs (MIR bands W1 and W2), and reduce the data using the method described in Section \ref{subsec:wise}. The reduced light curves are displayed in Figure \ref{fig:preflare}. Before this outburst, there is no significant variability except for a potential flare, which is only shown in the ASAS-SN light curve at MJD $\sim$ 56600 (in 2013), $\sim$970 rest-frame days before the first peak. Although the peak magnitude is comparable to the 2022 outburst, it is just above the detection limit of ASAS-SN, and it is not included in the ASAS-SN transient list\footnote{\url{https://www.astronomy.ohio-state.edu/asassn/transients.html}}. More importantly, no contemporary photometric or spectroscopic data can determine whether it is related to the recent nuclear outbursts or caused by a nearby supernova outburst, considering the large FWHM of $\sim$16'' for ASAS-SN \citep[][see also the ASAS-SN official website\footnote{\url{https://www.astronomy.ohio-state.edu/asassn/public}}]{Jayasinghe2018}. Therefore, we will not discuss this potential flare in the following text. For convenience, we refer to the flare that rose in 2022 as the ``first flare,'' and the flare that rises in 2024 as the ``second flare.''

\subsubsection{Light Curve Fitting}

\begin{figure*}[htb!]
    \plotone{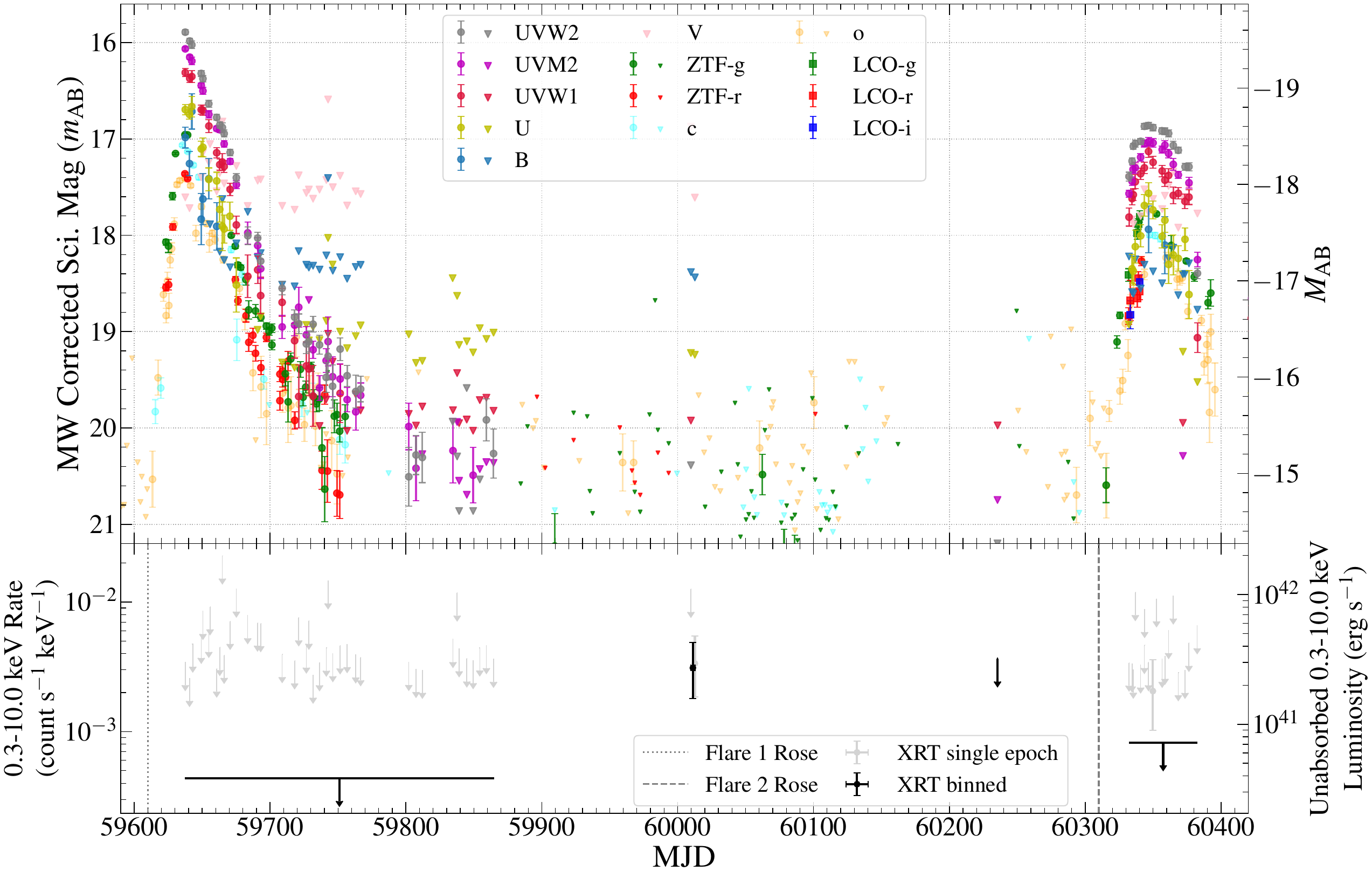}
    \caption{\textbf{Top panel:} The UV/optical light curves of AT\,2022dbl during the first and the second flare. 3-$\sigma$ upper limits are plotted in down triangles. \textbf{Bottom panel:} The X-ray count rate of AT\,2022dbl. The vertical dotted and dashed lines mark the approximate rise time of the first and second flares, respectively. 3-$\sigma$ upper limits are plotted in down arrows.}
    \label{fig:optuvx}
\end{figure*}

\begin{figure*}[htb!]
    \centering
    \epsscale{0.8}
    \plotone{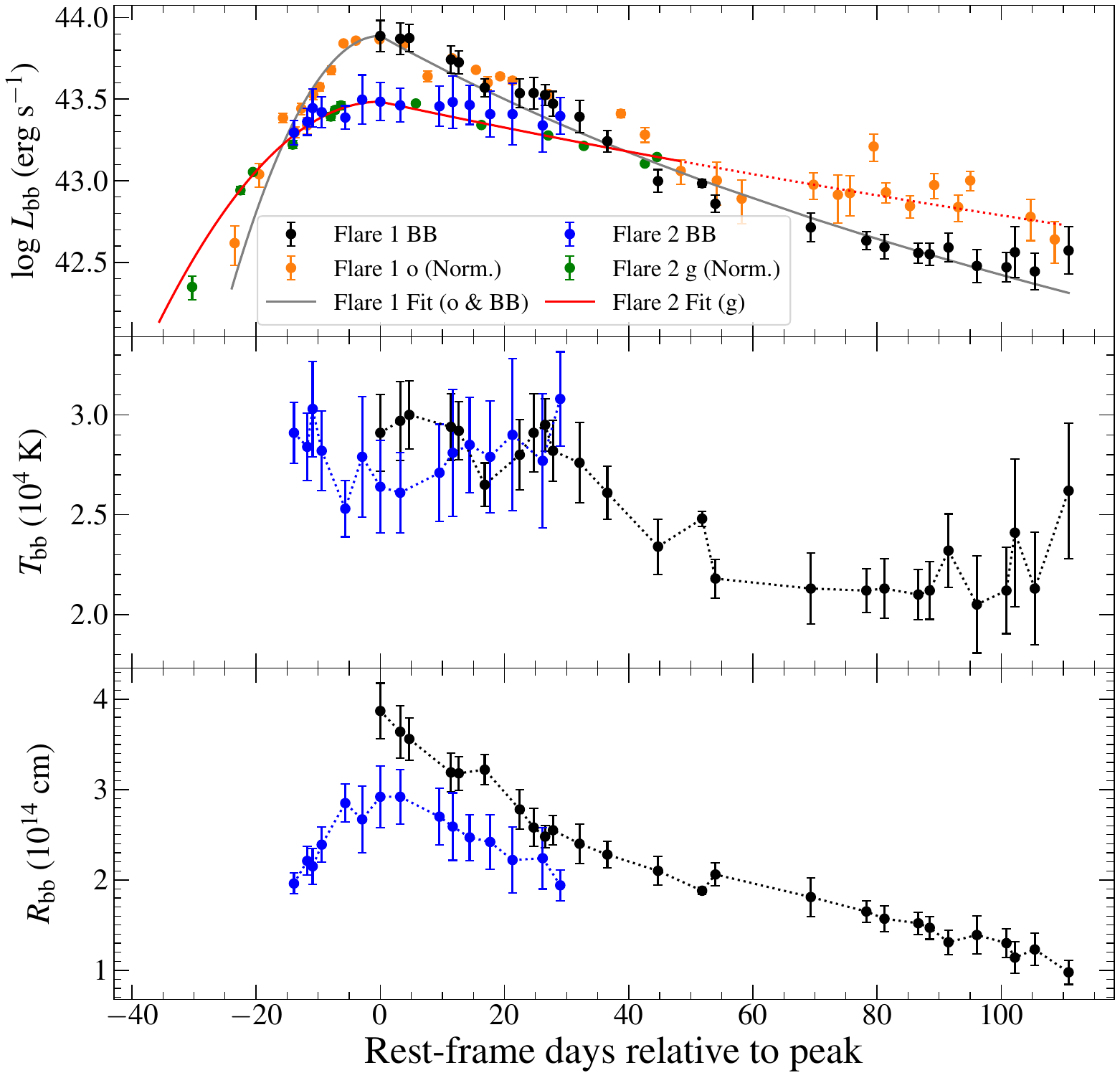}
    \caption{The light curve fitting results of AT\,2022dbl. The luminosities of the $o$- and $g$-bands are normalized to the blackbody luminosities of the first and second flares, respectively. For comparison, the fitted decline curve for the second flare is extended to $\sim$+120 d in the dotted style.}
    \label{fig:bb}
\end{figure*}

The optical/UV light curves during the first flare are displayed in the top panel of Figure \ref{fig:optuvx}. The rise stage of the first flare is well covered by the ATLAS $o$ band. Since the peak, the light curves are well covered by the Swift UVOT observations, and the first epoch happens to be around the peak. Therefore, we set the peak time to the first Swift epoch $t_{\rm peak1}$ = (MJD) 59637.6. For the UV/optical light curves since the peak, we use the \texttt{Superbol} package \citep{Nicholl2018} to interpolate the light curves and fit all photometry at each Swift epoch into a blackbody SED. The best-fit results are displayed in Figure \ref{fig:bb}. The blackbody temperature $T_{\rm bb}$ slowly declines from $\sim3\times10^4$ K to $\sim2\times10^4$ K. The blackbody radius $R_{\rm bb}$ smoothly declines from $\sim4\times10^{14}$ cm to $\sim1\times10^{14}$ cm. 


The optical/UV light curves during the second flare are displayed in the top panel of Figure \ref{fig:optuvx}. Its rise stage is well covered by the ZTF and LCO $g$ band. We choose the peak time as the brightest Swift epoch, $t_{\rm peak2}$ = (MJD) 60346.6, and perform the blackbody fitting on all photometry at Swift epochs except for the last one, which is apparently problematic. As shown in Figure \ref{fig:bb}, from $-$15 d to +30 d, the blackbody temperature remains fairly constant at $\sim$26000 K, while the blackbody luminosity evolves slowly, peaking $\sim$0.4 dex lower than the previous flare. Although the flat peak has been well covered by Swift, it has unfortunately entered safe mode since March 15, 2024, which was exactly when the source left the peak. After that, the decline stage is sparsely covered by the ZTF $g$ and ATLAS $o$ bands.





We characterize the light curves of both flares by the rest-frame rise time from half-peak luminosity to peak luminosity ($t_{\rm 1/2,rise}$) and the decline time from peak luminosity to half-peak luminosity ($t_{\rm 1/2,decline}$). To extract these two timescales, we fit the light curves with a Gaussian rise and a power-law decline:
\begin{align}\label{eqn:fitg}
    L(t)
    &= L(t_{\rm peak})\times
    \begin{cases}
        e^{-(t-t_{\rm peak})^2/(2\sigma^2)}, & t < t_{\text{peak}} ; \\
        \left(\frac{t-t_{\rm peak}+\tau}{\tau}\right)^{\alpha}, & t \geqslant t_{\text{peak}} .
    \end{cases}
\end{align}
For the first flare, the rise and decline fittings are performed on the $o$-band and blackbody luminosity, respectively. For the second flare, the fitting is performed on the $g$-band luminosity. The best-fit light curves are drawn in the top panel of Figure \ref{fig:bb}, and the fitted parameters are listed in Table \ref{tab:fit}.

\begin{deluxetable*}{ccccccc}
\caption{The best-fit light curve parameters for the two flares \label{tab:fit}}
\tablehead{
\colhead{Flare} & \colhead{$t_{\text{peak}}$} & \colhead{$L_{\text{BB,peak}}$} & \colhead{$T_{\text{BB,peak}}$} & \colhead{$R_{\text{BB,peak}}$} & \colhead{$t_{\rm 1/2,rise}$} & \colhead{$t_{\rm 1/2,decline}$} \\ 
\colhead{[No.]} & \colhead{[MJD]} & \colhead{[log (erg s$^{-1}$)]} & \colhead{[$10^4$ K]} & \colhead{[$10^{14}$ cm]} & \colhead{[day]} & \colhead{[day]} 
}
\startdata
1 & 59637.6 & 43.89 $\pm$ 0.10 & 2.91 $\pm$ 0.19 & 3.87 $\pm$ 0.31 & 10.6 $\pm$ 0.5 & 15.7 $\pm$ 0.8 \\
2 & 60346.6 & 43.48 $\pm$ 0.12 & 2.64 $\pm$ 0.23 & 2.92 $\pm$ 0.34 & 16.8 $\pm$ 0.5 & 36.9 $\pm$ 2.4 
\enddata
\tablecomments{$t_{\rm 1/2,rise}$: the rest-frame rise time from half-peak luminosity to peak luminosity. \\ $t_{\rm 1/2,decline}$: the rest-frame decline time from peak luminosity to half-peak luminosity.}
\end{deluxetable*}


\subsection{Optical Spectral Analysis} \label{subsec:specana}
As introduced in Section \ref{subsec:optspec}, 3 LCO spectra taken during the first flare were selected, while 4 optical spectra have been taken during the second flare. In addition, an SDSS host spectrum is available. All of these spectra are shown in Figure \ref{fig:optspec}. The spectral fitting procedures for each transient spectrum are listed as follows: 

\begin{figure*}[htb!]
    \centering
    \epsscale{0.85}
    \plotone{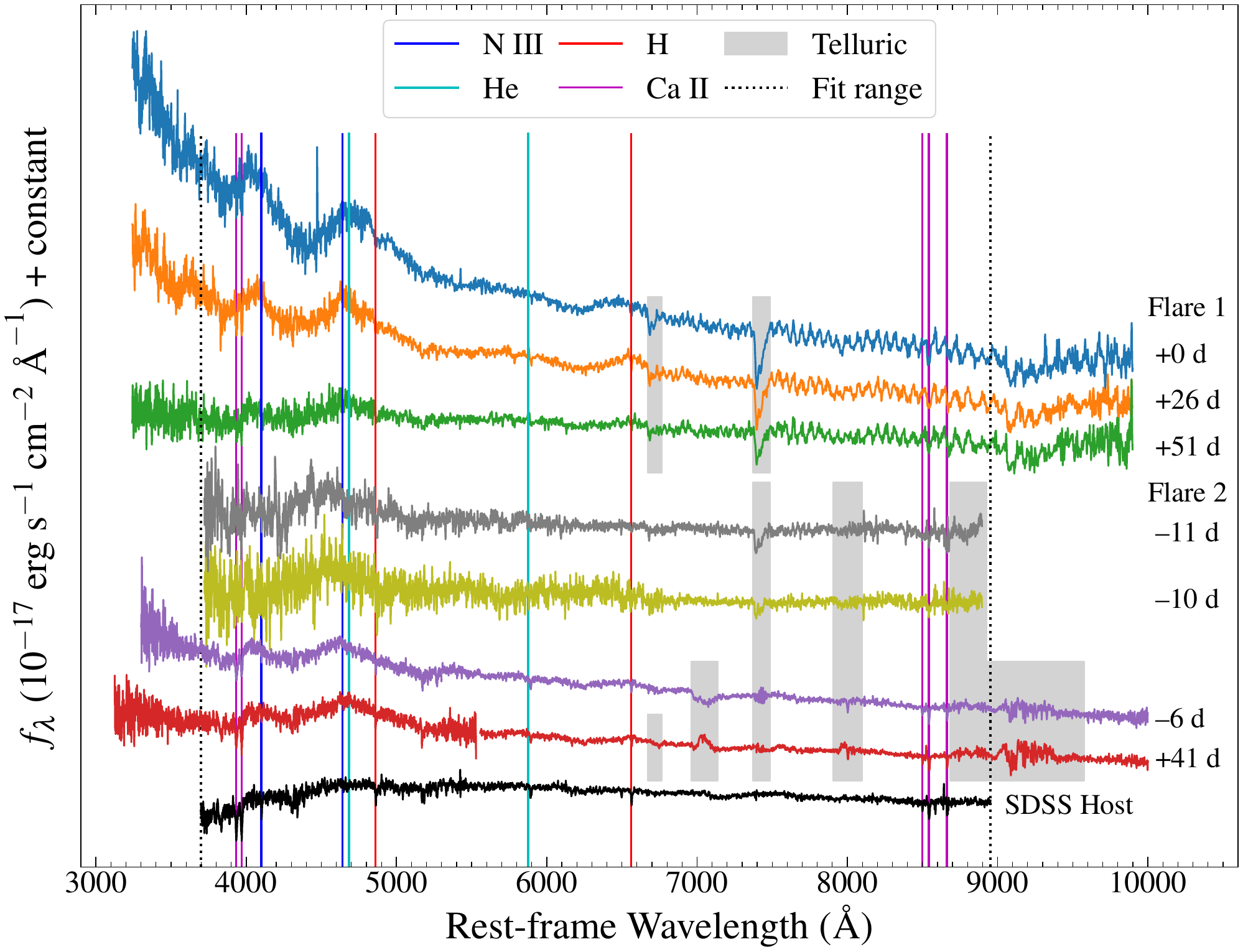}
    \caption{The optical spectra of AT\,2022dbl. Black dotted lines limit the range of the spectral fitting.}
    \label{fig:optspec}
\end{figure*}

(1) Host-galaxy subtraction. Since the host spectrum displays clear Ca \textsc{ii} absorption doublets at $3910-4000$ \AA, and the blue side has higher SNR than the red side, we used these doublets for calibration. We fit and subtract the nearby pseudo-continuum for both the transient and host spectra. Then a least-squares fitting on the residuals gives the multiplication factor for the host galaxy component. Limited by the wavelength range of the host spectrum, we perform the fitting only within this range. The three representative LCO spectra were taken at MJD 59638 (+0 d), MJD 59664 (+26 d) and MJD 59690 (+51 d). The two HCT spectra taken at the early stage of the second flare are discarded, as their SNRs are too low for the host-galaxy subtraction and also for further analysis.

(2) Continuum fitting. After subtracting the host component, a power-law function is used to fit the continuum. In the case of LCO spectra, the continuum windows are set to the following line-free regions (in rest-frame wavelengths): $3700-3900$ \AA, $5200-5400$ \AA, $6100-6300$ \AA, $7100-7400$ \AA\ and $7600-8490$ \AA, with the exclusion of the telluric absorption regions. For the P200 spectra, the continuum and telluric absorption regions are a bit different (see Figure \ref{fig:specfit1}). 

(3) Line fitting. After subtracting the continuum, all residuals exhibit multiple broad characteristics around $3900-4200$ \AA, $4400-5200$ \AA, and $6300-6900$ \AA, some showing a faint broad bump around $5500-6100$ \AA. The broad feature in the $3900-4200$ \AA\ range is symmetrical and peaks at approximately 4100 \AA, possibly corresponding to N \textsc{iii} (4100) or H$\delta$ (4101). In the range $4400-5200$ \AA, the characteristic is asymmetric and could be a combination of N \textsc{iii} (4640), He \textsc{ii} (4686) and H$\beta$ (4861). Lastly, the broad feature in the range $6300-6900$ \AA\ is symmetric and centers around 6560 \AA. It is consistent with a broad H$\alpha$ (6563) emission line; The $5500-6100$ \AA\ feature can be tentatively interpreted as He \textsc{i} (5876). The selection of the fitting components is based on these facts: First, the extended red wing of the $4400-5200$ \AA\ feature indicates the existence of H$\beta$, which is further supported by the existence of H$\alpha$. Second, it is unlikely that the $3900-4200$ \AA\ bump is dominated by H$\delta$, since H$\alpha$ is too weak compared to this feature. Therefore, it should be dominated by  N \textsc{iii} (4100), although the H$\delta$ will slightly affect the intensity. The N \textsc{iii} $\lambda$4100 lines are usually produced by the Bowen mechanism, which requires He \textsc{ii} Ly$\alpha$ lines at 304 \AA. Taking into account the extreme strength of N \textsc{iii} $\lambda$4100, the He \textsc{ii} emission should be strong. Moreover, the N \textsc{iii} $\lambda$4640 lines should also be produced via this mechanism. Therefore, both the He \textsc{ii} $\lambda$4686 line and the N \textsc{iii} $\lambda$4640 line should be considered. In addition, a He \textsc{i} $\lambda$5876 component is involved to cover the weak emission features in several spectra. To ensure reliability, the FWHM and the offset of the two features of N \textsc{iii} are tied up, as do those of H$\alpha$ and H$\beta$. The fitting results are shown in Figure \ref{fig:specfit1}. 

\begin{figure*}[htb!]
    \centering
    \epsscale{0.9}
    \plotone{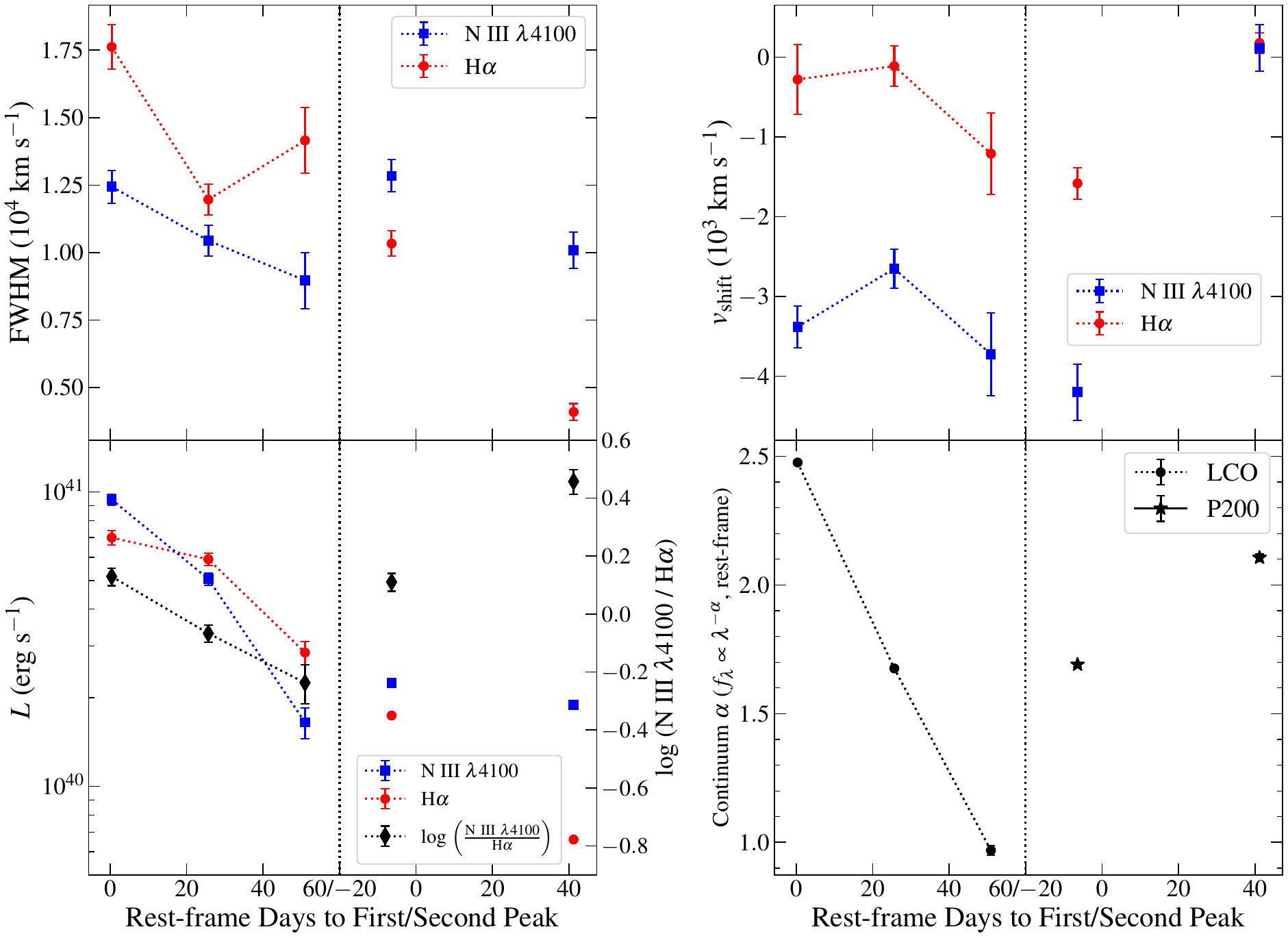}
    \caption{The evolution of the FWHM, velocity shift and luminosity of N \textsc{iii} $\lambda$4100 (blue squares) and H$\alpha$ (red dots) emission lines as well as their ratios (black diamonds), and the power-law indexes of the continua. In the last panel, measurements from the LCO spectra taken during the first flare and P200 spectra taken during the second flare are marked in dots and asterisks, respectively.}
    \label{fig:specevo}
\end{figure*}

Despite careful selection of fitting components, the $4400-5200$ \AA\ feature is still hard to deblend due to its smoothness, and hence it cannot prove or disprove the existence of He \textsc{ii} $\lambda4686$ and the associated Bowen mechanism as well as the intensity of N \textsc{iii} $\lambda4640$. Therefore, we only focus on the evolution of the most prominent and unblended features: N \textsc{iii} $\lambda$4100 and H$\alpha$. Furthermore, we also examine the power-law indexes of the continua. Figure \ref{fig:specevo} illustrates the evolution of the FWHM, velocity shift and luminosity of N \textsc{iii} $\lambda$4100 and H$\alpha$ emission lines, along with the power-law indexes of the continua, during both flares. 

For the LCO spectra taken during the first flare, the FWHMs for the H$\alpha$ lines in all spectra are well above 10000 km s$^{-1}$, showing a slowly narrowing trend from FWHM $\sim$18000 km~s$^{-1}$ to $\sim$12000 km~s$^{-1}$ during $+$0 d to $+$51 d to the first peak. N \textsc{iii} $\lambda$4100 shows a narrowing trend from FWHM $\sim$12000 km~s$^{-1}$ to $\sim$9000 km~s$^{-1}$. Except for the first epoch, neither of the N \textsc{iii} $\lambda$4100 lines nor H$\alpha$ exhibit clear shifts towards blue or red. The luminosity of H$\alpha$ and N \textsc{iii} $\lambda$4100 gets lower at later phases. The high N \textsc{iii} $\lambda$4100 luminosity of $\sim10^{41}$ erg s$^{-1}$ and the evolution of N \textsc{iii} $\lambda$4100 / H$\alpha$ ratio highly resemble AT\,2018dyb, which has the highest N \textsc{iii} $\lambda$4100 luminosity among spectroscopically confirmed TDEs \citep{Leloudas2019,Charalampopoulos2022}. We note that the precision of the luminosity depends on the flux calibration. The continuum gets flatter as it gets fainter. 

For the first P200 spectrum ($-$6 d to the second peak), it displays an H$\alpha$ emission line with FWHM $\sim$ 10000 km s$^{-1}$ and N \textsc{iii} $\lambda$4100 with FWHM $\sim$ 13000 km s$^{-1}$, while the velocity shift and luminosity for both lines are similar to the late-time spectra of the previous flare. 
The power-law index for the continuum rises again, which is consistent with a newly risen flare. The second P200 spectrum displays a much narrower and weaker H$\alpha$ feature with FWHM $\sim$ 4000 km s$^{-1}$ and a luminosity of $<10^{40}$ erg s$^{-1}$, which fades much quicker than the N \textsc{iii} $\lambda$4100 feature. As a result, the ratio N \textsc{iii} $\lambda$4100 / H$\alpha$ increases to $\gtrsim$1. The power-law index for the continuum is higher than that of the previous spectrum.

\subsection{X-ray Luminosity Estimation}\label{subsec:xray}
As described in Section \ref{subsec:swift}, X-ray observations were made with Swift XRT during both flares. All X-ray epochs are divided into four segments, as shown in the bottom panel of Figure \ref{fig:optuvx}. Only one segment yields a marginal detection with rate-to-error ratio of $\sim$2, which does not allow for spectral analysis (Bottom panel of Figure \ref{fig:optuvx}). The stacked image of the first phase (0 d to +220 d relative to the first peak) yields a total exposure time of 78.7 ks and a tight 3-$\sigma$ upper limit for the 0.3-10.0 keV count rate of $4.37\times10^{-4}$ counts s$^{-1}$. Assuming a typical blackbody model of $kT=50$ eV \citep[e.g.][]{Guolo2024X} and a hydrogen column density of $N_{\rm H}=1.94\times10^{20}\ {\rm cm}^{-2}$ \citep{HI4PI2016}, we obtain an upper limit for the unabsorbed luminosity using the WebPIMMS tool\footnote{\url{https://heasarc.gsfc.nasa.gov/cgi-bin/Tools/w3pimms/w3pimms.pl}} of $L_{{\rm X},1}<3.8\times10^{40}$ erg~s$^{-1}$. Following the same method, we derive the luminosity for the later three segments: $L_{{\rm X},2}=2.7^{+1.5}_{-1.1}\times10^{41}$ erg s$^{-1}$, $L_{{\rm X},3}<3.2\times10^{41}$ erg s$^{-1}$, $L_{{\rm X},4}<7.2\times10^{40}$ erg s$^{-1}$.


\section{Discussion} \label{sec:discuss}
\subsection{AT\,2022dbl as a robust repeated pTDE}
We shall discuss the origin of these two flares as follows. As displayed in Section \ref{subsec:hostana}, the pre-outburst SDSS spectrum exhibits a series of strong Balmer absorption lines and shows no clear emission line after subtracting the best-fit stellar continuum. Combined with the lack of strong historical radio, X-ray and MIR variability, as well as the MIR W1$-$W2 color of 0.007 that against the AGN criterion, the presence of a persistent AGN can be firmly excluded. In addition, the first flare lasted less than a year, which is unusual for a turn-on AGN, and the second flare showed a number of similar photometric and spectroscopic features. We thereby reject the possibility of an AGN origin for both flares. On the other hand, both flares show broad H$\alpha$ emission with FWHM $>$ 10000 km s$^{-1}$ and declining blackbody radii after the peak, which also strongly contradicts the SN origin. 

\begin{figure*}[htb!]
    \centering
    \epsscale{1.}
    \plotone{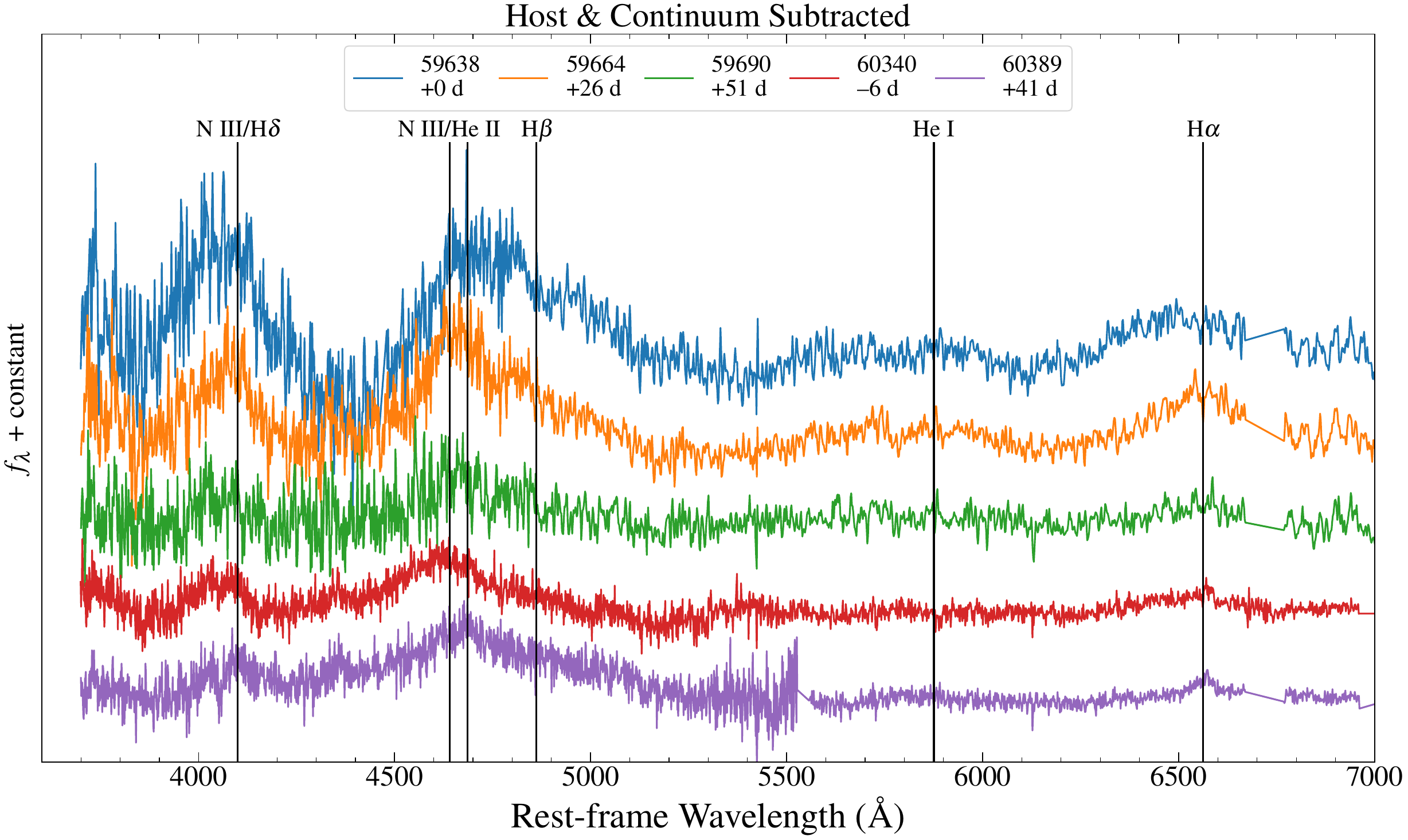}
    \caption{Spectra after subtracting the host galaxy contribution and the continuum. Spectra of both flares show similar emission lines.}
    \label{fig:specsimilar}
\end{figure*}

All of the features that disfavor AGNs and supernovae are nevertheless characteristic of TDEs, including the timescales of both flares, the fairly steady blackbody temperatures of $(2-3)\times10^4$ K, the value and evolution of the blackbody radii, and the very broad H$\alpha$ emission. Therefore, AT\,2022dbl is undoubtedly a repeated TDE. Moreover, both flares display highly similar broad H$\alpha$, $\sim$4400$-$5200 \AA\ (H$\beta$ \& possible N \textsc{iii} and He \textsc{ii}) and $\sim$4100 \AA\ (N \textsc{iii} \& possible H$\delta$) features, as shown in Figure \ref{fig:specsimilar}. In particular, for both flares, the luminosity of $\sim$4100 \AA\ is comparable to that of H$\alpha$ (See the lower left panel of Figure \ref{fig:specevo}), which is rare among all TDEs. Hence, these two flares are probably originated from the debris of a single disrupted star, and AT\,2022dbl should be a robust repeated pTDE.

We now try to rebuild the orbit of this "unluckiest star", before it got stripped by the BH. Assuming a BH mass of $10^{6.40}\ M_\odot$, and an elliptical orbit with a period of $\sim$710 days, the semi-major axis of the orbit should be log $a$ (cm) = 15.5, or $a\approx210$ AU. For a solar-like star, the tidal radius should be log $R_{\rm t}$ (cm) = 13.0, or $R_{\rm t}\approx0.6$ AU. Hence, the eccentricity is $e=1-R_{\rm p}/a\sim 1-R_{\rm t}/a\approx 0.997$. Based on this result, we tentatively propose a scenario for repeated pTDEs at the end of Section \ref{subsec:robust}.

\subsection{Comparison with other repeated pTDEs and common optical TDEs}
As mentioned in Section \ref{sec:intro}, several repeated pTDE candidates have been reported in the literature: IC 3599, ASASSN-14ko, eRASSt J045650.3–203750, Swift J023017.0+283603, AT\,2018fyk, RX J133157.6–324319.7 and AT\,2020vdq. We briefly list the information of these repeated pTDEs in Table \ref{tab:pTDE}. 

\begin{deluxetable*}{cccccc}
\caption{List of published repeated pTDE candidates\label{tab:pTDE}}
\tablehead{
\colhead{Name} & \colhead{Host Type} & \colhead{Band} & \colhead{Period/Interval (Days)} & \colhead{Flares} & \colhead{Peak Evolution}
}
\startdata
ASASSN-14ko$^{1,2,3,4}$ & Seyfert 2 & Opt./UV/X-ray$^\dagger$ & 115.2 & $\sim$30 & Similar \\
Swift J023017.0+283603$^{5,6}$ & Weak AGN & X-ray & $\sim$22 & $\sim$11 & Variable\\
eRASSt J045650.3–203750$^{7,8}$ & Quiescent & X-ray/UV$^\dagger$ & 299$\to$193 & 5 & Lower\\
\hline
IC 3599$^{9,10,11,12,13}$ & Seyfert 1.9 & X-ray/Opt.* & $\sim$3470$^?$ & 2/3 & Similar \\
AT\,2018fyk$^{14,15,16}$ & Quiescent & UV/X-ray & $\sim$1200 & 2 & Lower \\
RX J133157.6-324319.7$^{17,18}$ & Quiescent & X-ray & $\sim$10000 & 2 & Similar\\
AT\,2020vdq$^{19,20,21}$ & E+A & Opt./UV*/X-ray* & $\sim$870 & 2 & Higher\\
AT\,2022dbl$^{22}$ & QBS & Opt./UV & $\sim$710 & 2 & Lower\\
\enddata
\tablecomments{\\-- Band: $^\dagger$ Not periodic. * Not observed during the first flare. \\ 
-- Period/Interval: ASASSN-14ko shows a nearly constant period of 115.2 days. Swift J023017.0+283603 shows a period of $\sim$22 days. eRASSt J045650.3–203750 has shown 5 flares with the interval declining from 299 days to $\sim$193 days. Other sources show only two flares. $^?$ IC 3599 has showed two prominent X-ray flares in 1990 and 2010. \citet{Campana2015} predicted a 9.5 yr period in a repeated pTDE scenario, suggesting a missing flare between the two flares. However, \citet{Grupe2024} reported that another X-ray flare did not come in the predicted time window. \\-- Peak Evolution: The peak luminosity of the earlier flare versus that of the later flare. \\-- References: 1. \citet{Payne2021}, 2. \citet{Payne2022}, 3. \citet{Payne2023}, 4. \citet{Huang2023}, 5. \citet{Evans2023}, 6. \citet{Guolo2024}, 7. \citet{Liu2023}, 8. \citet{Liu2024}, 9. \citet{Grupe1995}, 10. \citet{Komossa1999}, 11. \citet{Grupe2015}, 12. \citet{Campana2015}, 13. \citet{Grupe2024}, 14. \citet{Wevers2019}, 15. \citet{Wevers2023}, 16. \citet{Pasham2024}, 17. \citet{Hampel2022}, 18. \citet{Malyali2023}, 19. \citet{Yao2023}, 20. \citet{Som2023}, 21. \citet{Somalwar2023}, 22. This work.}
\end{deluxetable*}

\begin{figure*}[htb!]
    \gridline{\fig{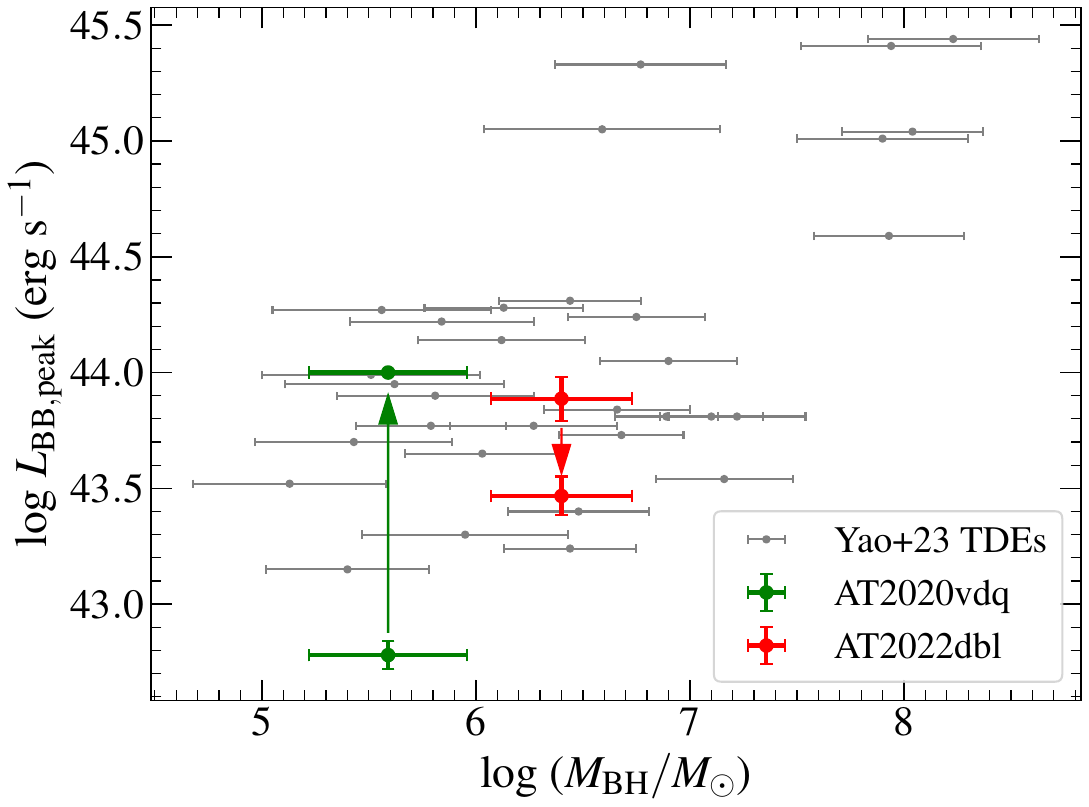}{0.4\textwidth}{(a)}
               \fig{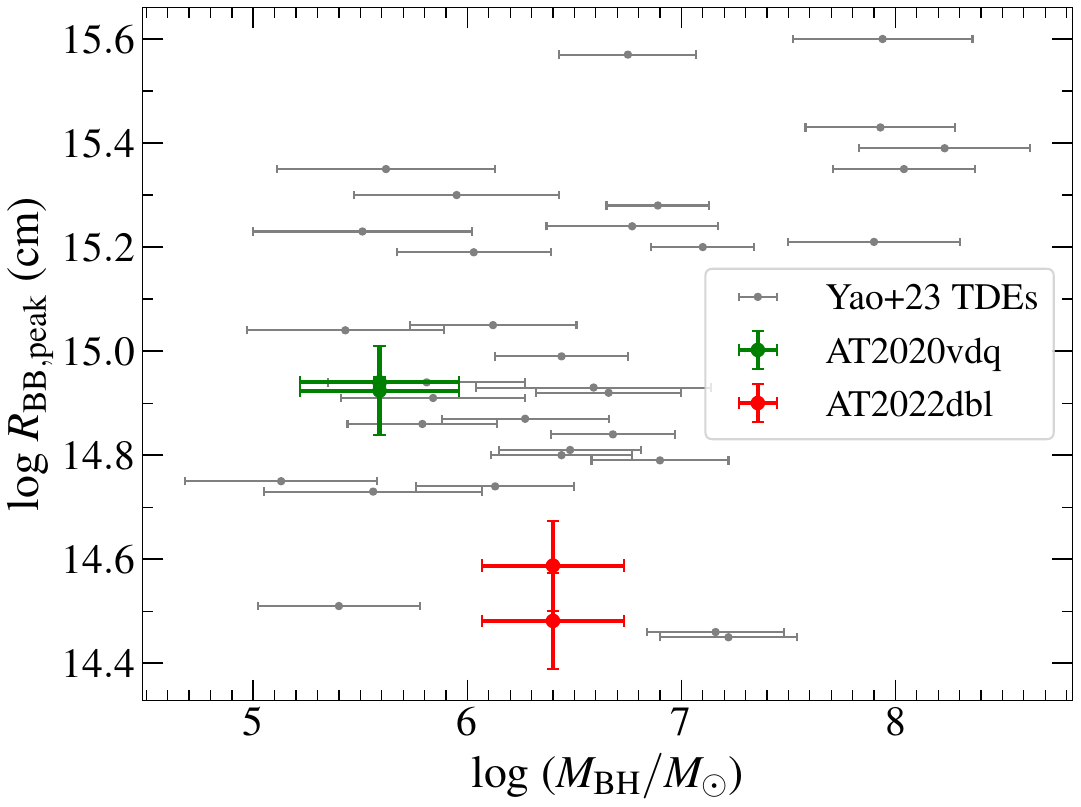}{0.4\textwidth}{(b)}}
    \gridline{\fig{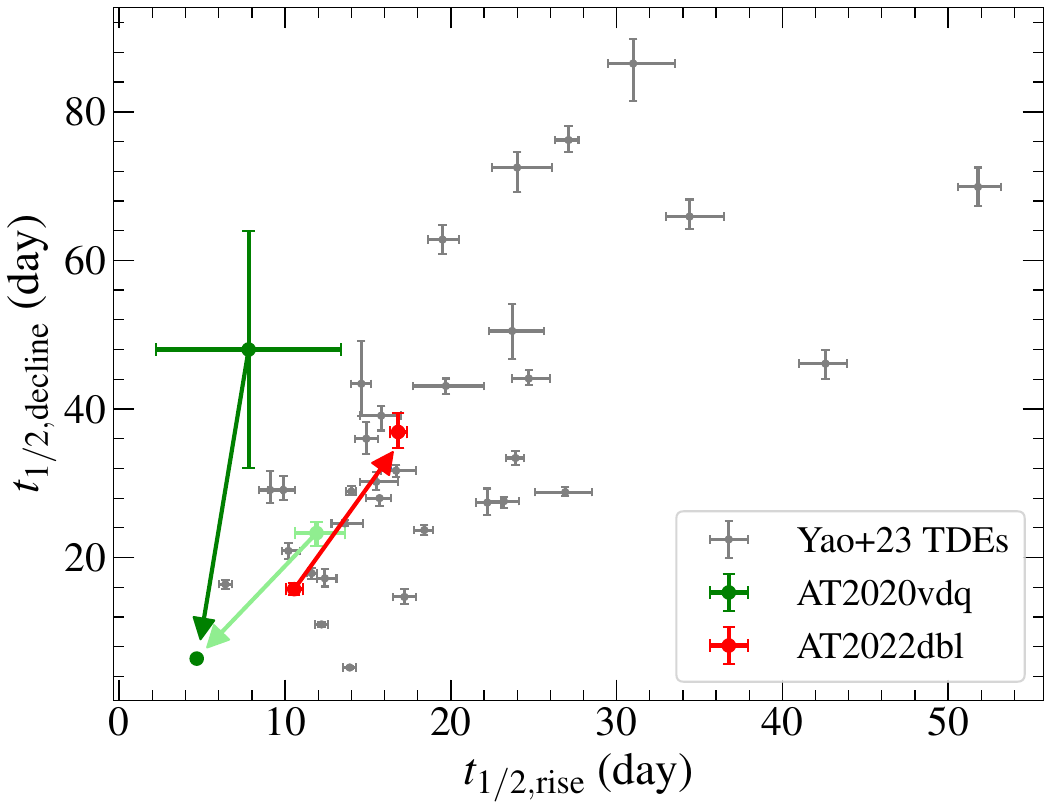}{0.4\textwidth}{(c)}}
              
    \caption{Comparison of optical repeated pTDEs AT\,2020vdq \citep{Yao2023,Somalwar2023} and AT\,2022dbl (This work), as well as optical TDEs listed in \citet{Yao2023}. Black hole mass versus (a): Peak blackbody luminosity; (b): Peak blackbody radius; (c): Rest-frame rise time from half-peak luminosity to peak luminosity versus decline time from peak luminosity to half-peak luminosity. \\
    Note: (1) In Plot (b), both AT\,2020vdq and AT\,2022dbl show similar blackbody radius in their two flares. (2) The parameters of AT\,2020vdq are mostly adopted or derived from \citet{Somalwar2023}. However, in Plot (c), the derived rise and decline timescales of the first flare of AT\,2020vdq in \citet{Yao2023} and \citet{Somalwar2023} are greatly different, hence plotted in light-green and green, respectively.}
    \label{fig:compTDEs}
\end{figure*}

As shown in the table, only ASASSN-14ko, AT\,2020vdq and AT\,2022dbl show recurring flares in optical bands. We only focus on the comparison of AT\,2020vdq and AT\,2022dbl, as they share similar intervals and host galaxy types, while the behavior of ASASSN-14ko differs greatly. We compare the peak blackbody luminosity and radius, the rise and decline timescales of AT\,2022dbl with those of AT\,2020vdq and other ZTF TDEs listed in \citet{Yao2023}, as plotted in Figure \ref{fig:compTDEs}.

AT\,2022dbl shows several differences compared to AT\,2020vdq. First, its peak luminosity for the second flare is $\sim$0.4 dex lower than that of the first flare; While for AT\,2020vdq, the peak luminosity of the second flare is $\sim$1.2 dex higher than that of the first flare. Second, for AT\,2022dbl, the second flare rises and declines slower than the first flare. In contrast, for AT\,2020vdq, the second flare rises and declines much more quickly than the first flare. 

We compare AT\,2020vdq and AT\,2022dbl with the ZTF TDEs \citep{Yao2023} that show no recurrent flare by now. For AT\,2020vdq, the peak luminosity of its first flare are the lowest and the second lowest among all optical TDEs, while the rise and decline timescales of its second flare are the lowest and the second lowest among all optical TDEs. In short, both flares of AT\,2020vdq show some peculiarities compared to normal TDEs. For AT\,2022dbl, its peak blackbody radius is the smallest among all TDEs with 6 $\leqslant$ log $(M_{\rm BH}/M_\odot)$ $\leqslant$ 7. Apart from this, its peak luminosity, rise and decline timescales of both flares are all typical among the ZTF TDEs. Therefore, the two flares of AT\,2022dbl are both typical tidal disruption flares.

\subsection{Robustness of a "repeated pTDE" classification}\label{subsec:robust}
As introduced in Section \ref{sec:intro}, the identification of a repeated pTDE can be complicated by some alternative origins. Hence, a robust identification of a repeated pTDE (especially an optical one) is difficult. It requires not only confirmation of the TDE origin but also a trustworthy connection between the flares. 

As an example, we examine the case of AT\,2020vdq. In \citet{Som2023}, the authors establish the TDE origin for the first flare by the broadband light curve, the newly risen radio flare, the E+A host galaxy, as well as the intermediate width ($\sim$700$-$1000 km s$^{-1}$) Balmer, He \textsc{ii}, He \textsc{i}, and [Fe \textsc{x}] emission in the late-time spectra ($\sim$+600 d). In \citet{Somalwar2023}, the second flare is spectroscopically identified as a TDE, since the spectra around the peak exhibit broad ($\sim$20000 km s$^{-1}$) Balmer, He \textsc{ii} and He \textsc{i} emission lines. Although the TDE-H+He identification for both flares is reliable, the two flares show highly different peak luminosity and light curve shapes and have no contemporary spectra to support their physical connection. Moreover, the E+A host galaxy may have a much higher TDE rate than normal galaxies \citep{Arcavi2014,French2016,Hammerstein2021}. In an extreme case, the probability of detecting two independent TDEs within $\sim$3 years can be as high as 30\% \citep[See Section 5.1 of][]{Somalwar2023}. 

In the case of AT\,2022dbl, its two flares not only exhibit photometric and spectroscopic features that firmly establish their TDE origins, but also display similar broad Balmer, N \textsc{iii}, and possible He \textsc{ii} emission lines in the early spectra of both flares, strongly indicating a connection between them (see Figure \ref{fig:specsimilar}). This represents the first robust spectroscopic evidence for a repeated pTDE. 

This spectroscopic evidence is important for the repeated pTDE classification, as current photometric data for both events cannot provide enough support. On the one hand, there are only two flares in both events, allowing for alternative origins, especially independent TDEs, as their host galaxies can have higher TDE rates than normal galaxies. On the other hand, the light curves can provide limited information on the judgment of repeated pTDEs by now, as there is currently lack of reliable optical/UV repeated pTDEs for comparison. A third flare can provide the conclusive evidence for a repeated pTDE classification, which might occur in the next couple of years.

We notice that recent simulation works of repeated pTDEs \citep[e.g.][]{Bandopadhyay2024,LiuC2024} have reproduced the light curve patterns similar as ASASSN-14ko, AT\,2020vdq and AT\,2022dbl, under several sets of stellar parameters (e.g., mass and age) and $\beta$. These results are heuristic but still preliminary, as the parameter space remains to be fully explored. If the simulation sets can be extended to a grid, we can constrain the stellar parameters and $\beta$, and predict the future evolution.

As illustrated in Figure \ref{fig:compTDEs}(b), for both events, their two flares share similar peak blackbody radius. The radius for AT\,2022dbl is log $R_{\rm bb}$ (cm) $\sim14.5$, which lies between the tidal radius, log $R_{\rm t}$ (cm) = 13.0 and the semi-major axis, log $a$ (cm) = 15.5. This relation is also found in AT\,2020vdq: log $R_{\rm bb}$ (cm) $\sim14.9$, log $R_{\rm t}$ (cm) $=12.7$, log $a$ (cm) $=15.3$. This similarity provides additional support for a pTDE claim, as it suggests the connection between these two flares. Based on the relation, we tentatively propose this scenario: The star shallowly encounters the SMBH and loses a small fraction of mass, then leaves the tidal radius with the orbit largely unaffected. As a result, the bound debris can self-intersect at a similar radius, which is far away from both the pericenter and apocenter. Additional theoretical works and numerical simulations are encouraged to test this scenario.

\section{Conclusion} \label{sec:conclude}
We have reported the discovery of a repeated partial TDE AT\,2022dbl in a nearby quiescent galaxy. In this event, two separate flares occurred in 2022 and 2024, with an interval of $\sim$710 days. Both flares have been fortunately followed by high-cadence optical/UV photometry and X-ray observations, as well as a series of optical spectroscopy observations, which help to confirm the TDE origin for both flares. More importantly, similar broad Balmer, N \textsc{iii} and possible He \textsc{ii} emission lines, especially the extreme $\sim$4100 \AA\ emission lines, help to rule out the possibility of two independent TDEs and provide the first robust spectroscopic evidence for two tidal disruptions of the same star. 

Both flares of AT\,2022dbl are bright in optical/UV wavelengths but much fainter in X-ray, which are similar to most TDEs that found in optical surveys in the past decade. Repeated pTDEs, particularly optical/UV bright TDEs like AT\,2022dbl, provide valuable opportunities to test optical/UV emission models, as another flare is expected in the coming years. Its repeatability enables us to carefully plan for multi-wavelength observations of subsequent flares from the earliest stages. With the assistance of high-cadence optical/UV/X-ray photometric and spectroscopic data, we can take the chance to collect important clues to the mechanism of optical/UV emission of TDEs, as well as the associated "missing energy" problem \citep{Piran2015,Lu2018}. As the next-generation "TDE hunters" come into play, such as the Vera Rubin Observatory \citep[VRO;][]{LSST2019} and the Wide Field Survey Telescope \citep[WFST;][]{Lin2022,WFST2023}, the high-cadence multiband surveys are expected to reveal a number of such pTDEs and accelerate the process of solving these puzzles in the near future. \\




This work is supported by the National Natural Science Foundation of China (12233008, 12393814, 12073025, 12192221), the National Key R\&D Program of China (2023YFA1608100), the Strategic Priority Research Program of the Chinese Academy of Sciences (XDB0550200, XDB41000000), the China Manned Space Project (CMS-CSST-2021-A13, CMS-CSST-2021-A07), the Cyrus Chun Ying Tang Foundations, the Fundamental Research Funds for Central Universities (WK3440000006) and the Anhui Provincial Natural Science Foundation (2308085QA32). 
K.M. acknowledges support from JSPS KAKENHI grant No. JP24H01810. The authors appreciate the support of the Cyrus Chun Ying Tang Foundations. 
We thank the Swift science operations team for accepting our ToO requests and arranging the observations. We thank all researchers who have submitted Swift ToO requests and LCO spectroscopy proposals. We thank the staff of IAO, Hanle, CREST, and Hosakote, who made these observations possible. The facilities at IAO and CREST are operated by the Indian Institute of Astrophysics, Bangalore. We thank Prof. Christoffer Fremling and Nicholas Earley for helping us obtain a P200 spectrum on March 20, 2024. We acknowledge the support of the staff of the Lijiang 2.4m telescope, although we failed to obtain a usable spectrum. Z.L. thanks Dr. Junbo Zhang, Dr. Jie Zheng and Dr. Junjie Jin for the kind help on the ToO observation on the Xinglong 2.16m telescope and the subsequent data reduction, and sincerely apologizes for not using this low-resolution spectrum. Z.L. sincerely thanks the UK Swift Science Data Centre (UKSSDC) helpdesk (especially Phil and Kim) for the kind instructions and \textit{Swift} replies on the reduction of XRT data, and thanks Robert Wiegand for the help on the reduction of UVOT data.  The ZTF forced-photometry service was funded under the Heising-Simons Foundation grant \#12540303 (PI: Graham). This research
uses data obtained through the Telescope Access Program
(TAP). Observations with the Hale Telescope at Palomar Observatory were obtained as part of an agreement between the
National Astronomical Observatories, Chinese Academy of
Sciences, and the California Institute of Technology.

%

\vspace{5mm}





\appendix
\section{Observation \& Data Reduction} \label{sec:obs}

\subsection{ZTF Optical Photometry} \label{subsec:ztf}
The ZTF differential point-spread-function (PSF) photometry of AT\,2022dbl is obtained through the ZTF Forced-Photometry Service \citep{Masci2019}. We clean the photometry results by filtering out epochs that are impacted by bad pixels, and requiring thresholds for the signal-to-noise ratio of the observations, seeing, zeropoint, the sigma-per-pixel in the input science image, and the 1-$\sigma$ uncertainty on the difference image photometry measurement. We perform the baseline correction by the following two steps. First, we classify the measurements by the field, charge-coupled device (CCD) and quadrant identifiers. Then, for each class, we set the median of pre-outburst counts as the offset. After that, we build the ZTF $g$- and $r$- band light curves for AT\,2022dbl. AT\,2022dbl was first alerted by ZTF in March 2018 and got the internal name ZTF18aabdajx, and reported to TNS as AT\,2018mac. However, we carefully examine the light curves and confirm a false alert, which may be a temporary problem during the early test of ZTF.

\subsection{ATLAS \& ASAS-SN Optical Photometry}\label{subsec:atlasasas}
We obtain the ATLAS differential photometry from the ATLAS forced photometry server \citep{Tonry2018,Smith2020,Shingles2021}. To improve the signal-to-noise ratio (SNR), we combine the data into 1-day bins and build the ATLAS $c$- and $o$-band light curves. Meanwhile, we obtain ASAS-SN differential photometry from the ASAS-SN sky patrol \citep{Shappee2014,Kochanek2017}. The Galactic extinction corrected light curves are shown in the top panel of Figure \ref{fig:optuvx}.

\subsection{LCO Optical Photometry} \label{subsec:lco}
From January 22, 2024 to January 31, 2024, we conducted optical monitoring using the Las Cumbres Observatory Global Telescope network \citep[LCOGT;][]{LCOGT} in the $u$-, $g$-, $r$- and $i$-band with daily cadence.  With the same method of \cite{Zhu2023}, we use PanSTARRS (\citealt{PS1}) $gri$ band stack images as reference images and employ {\tt HOTPANTS}\,\citep{Becker2015} for image subtraction. After image subtraction, we perform PSF photometry on the difference image, and the photometric results are calibrated using PS1 standards in the field of view.
The Galactic extinction corrected photometric measurements are plotted in the top panel of Figure \ref{fig:optuvx}.

\subsection{Gaia, CRTS \& PTF Optical Photometry}\label{subsec:arch}
To check historical variability, we query the Gaia Photometric Science Alerts and the Catalina Real-Time Transient Survey (CRTS) \citep{Drake2009} and Palomar Transient Facility (PTF) catalogs. To improve SNR, we combine the CRTS and PTF data into 10-day bins. The results are displayed and discussed in Section \ref{subsec:preflare}.

\subsection{Swift UVOT \& XRT Observations} \label{subsec:swift}
The previous flare was fortunately well covered by Swift observations. During the previous flare, observations were performed by the X-Ray Telescope \citep[XRT;][]{Burrows2005} and the Ultra-Violet/Optical Telescope \citep[UVOT;][]{Roming2005} on Swift under a great number of ToO requests (Obs. ID: 00015026001-00015026045; PIs: Arcavi/Hinkle/Jiang/Makrygianni/Holoien/Margutti).
The recent flare had been well followed under several ToO requests (Obs. ID: 00015026046-00015026064; PIs: Lin/Hammerstein), before Swift unfortunately entered the safe mode on March 15, 2024.
We retrieve the Swift data from HEASARC\footnote{\url{https://heasarc.gsfc.nasa.gov/cgi-bin/W3Browse/swift.pl}} and process all data with \texttt{heasoft} v6.30.1. Details are described below.

\begin{figure}[htb!]
    \centering
    \figurenum{A1}
    \epsscale{0.7}
    \plotone{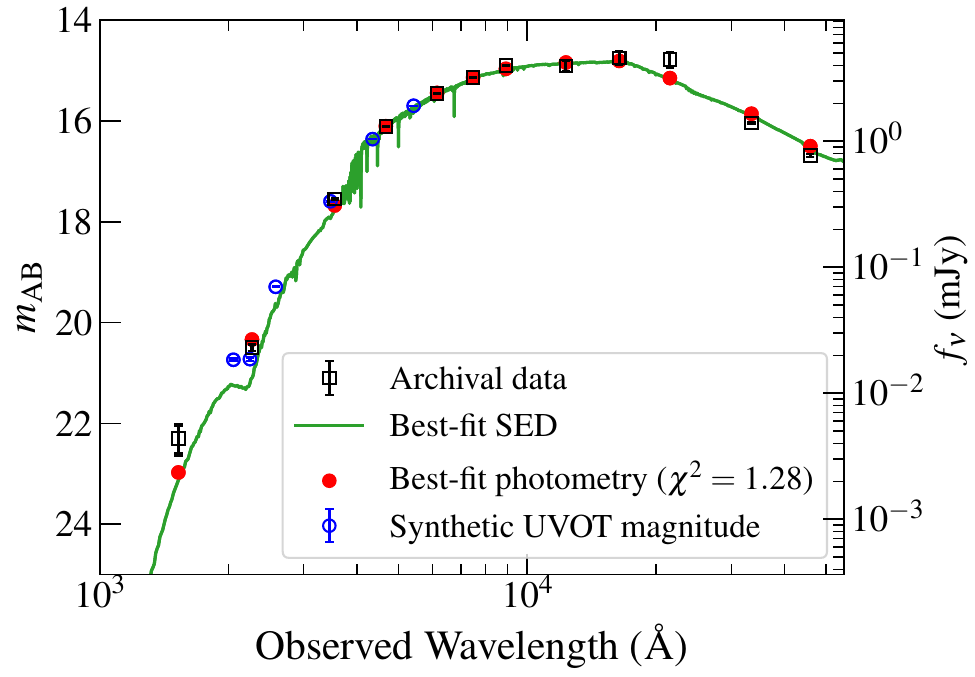}
    \caption{The SED fitting result of the \texttt{prospector} package, in the unit of AB magnitudes. The best-fit SED and photometry are plotted in green and red, respectively. The blue circles represent the synthetic magnitudes of UVOT UVW2, UVM2, UVW1, U, B and V bands.}
    \label{fig:sedfit}
\end{figure}

For each UVOT epoch, we first examine each image file and exclude the extensions with bad photometric flags. For image files with multiple valid extensions, we sum all extensions using the task \texttt{uvotimsum}. Then, the task \texttt{uvotsource} performs photometry on each image, with the source and source-free background region defined by a circle of the radius of $20^{\prime\prime}$ and $40^{\prime\prime}$, respectively. 
The host contribution is estimated via SED fitting. We collect the photometry from the Galaxy Evolution Explorer \citep[GALEX;][]{Martin2005} General Release 6, the Sloan Digital Sky Survey Data Release 16 \citep[SDSS DR16;][]{SDSSDR16}, the Two Micron All-Sky Survey \citep[2MASS;][]{Skrutskie2006}, and the AllWISE catalog \citep{Cutri2014}, listed in Table \ref{tab:galphoto}. After correcting the Galactic extinction, we fit the SED by running a dynamic nested sampler \texttt{dynesty} \citep{Speagle2020} under the \texttt{prospector} package \citep{Johnson2021}. The best-fit SED and synthetic UVOT magnitudes are plotted in Figure \ref{fig:sedfit}. The stellar mass derived from the posterior distribution is $\sim$$10^{10.47}\,M_{\odot}$, corresponding to a black hole mass of $\sim$$10^{6.89\pm0.26}\,M_{\odot}$ via the \citet{Reines2015} method.

\begin{deluxetable}{cccc}
\tablenum{A1}
\caption{Host-galaxy photometry used in the SED fitting \label{tab:galphoto}}
\tablehead{
\colhead{Catalog} & \colhead{Band} & \colhead{$\lambda_{\rm eff}$ (nm)} & \colhead{Flux (mJy)}}
\startdata
GALEX & FUV & 153 & 0.004 $\pm$ 0.001 \\
GALEX & NUV & 227 & 0.023 $\pm$ 0.001 \\
SDSS & $u$ & 355 & 0.346 $\pm$ 0.006 \\
SDSS & $g$ & 467 & 1.311 $\pm$ 0.004 \\
SDSS & $r$ & 616 & 2.388 $\pm$ 0.007 \\
SDSS & $i$ & 747 & 3.207 $\pm$ 0.009 \\
SDSS & $z$ & 892 & 3.994 $\pm$ 0.023 \\
2MASS & $J$ & 1232 & 3.947 $\pm$ 0.386 \\
2MASS & $H$ & 1642 & 4.554 $\pm$ 0.566 \\
2MASS & $K_s$ & 2157 & 4.432 $\pm$ 0.624 \\
WISE & W1 & 3346 & 1.394 $\pm$ 0.033 \\
WISE & W2 & 4595 & 0.771 $\pm$ 0.021 
\enddata
\end{deluxetable}

For each XRT epoch, we reduce the data by \texttt{xrtpipeline} and obtain the level 2 products. Then we use \texttt{xrtproducts} to extract the level 3 products. After that, we use \texttt{xselect} to stack all images. On this stacked image, no discernible source is shown at the position of the transient. To obtain an upper limit, the source region is selected as a circle of radius 20$\arcsec$, while the background region is defined as a source-free annulus with an inner radius of 50$\arcsec$ and an outer radius of 150$\arcsec$. We obtain the source and background photon counts in 0.3-10 keV by \texttt{ximage}. For images with photon counts in the source region $N\leqslant80$, a Bayesian approach is applied to calculate the 3-$\sigma$ lower and upper limits \citep{Kraft1991}; While for $N>80$, a Gaussian approach is adopted \citep{Evans2007,Evans2009,Konig2022}. Based on single-epoch photometric results, we divide all epochs into four segments: (1) Epochs before MJD 59900. All of the results are upper limits, so we stack all images to get a tighter upper limit. The total exposure time is 78.7 ks. (2) Epochs between MJD 60000 and 60030. The two epochs are isolated from the others; one of them yields a tentative detection with rate-to-error ratio of $\sim$ 2. The total exposure time is only 2.3 ks. (3) One epoch on MJD 60235. It is just before the rise of the second flare (MJD $\sim$ 60310), with an exposure time of 4.15 ks. (4) Since MJD 60310. Only one epoch reveals tentative detection with rate-to-error ratio of $\sim$ 2, and the total exposure time is 34.7 ks. The X-ray light curve is displayed in the bottom panel of Figure \ref{fig:optuvx}. The low SNR impedes us from further analysis, a brief luminosity estimation is introduced in Section \ref{subsec:xray}.

\subsection{Optical Spectroscopy} \label{subsec:optspec}
Since the discovery of the recurrent flare, we have obtained two spectra using the Double Spectrograph \citep[DBSP;][]{Oke1982} on the 200 inch Hale telescope at the Palomar Observatory (P200), and two spectra using the Himalaya Faint Object Spectrograph (HFOSC) instrument mounted on the 2-m Himalayan Chandra Telescope (HCT) of the Indian Astronomical Observatory \citep[IAO,][]{2014Prabhu}. The spectroscopic data are reduced in a standard manner using the packages in \texttt{IRAF} with the aid of the Python scripts hosted at \textsc{RedPipe} \citep{2021redpipe}.
We use \texttt{pypeit} package \citep{Prochaska2020} to reduce the P200/DBSP spectra, and extract the HCT
spectra by \texttt{IRAF}. As we retrieve the LCO photometric data, we find 12 automatically reduced public spectra taken by the 2.0m telescope at Haleakala Observatory, during the previous flare (Proposals: CON2022A-007/HAW2022A-002)\footnote{These spectra can be retrieved from the LCO Science Archive: \url{https://archive.lco.global}}. We use 3 high-quality representative spectra of them, which are introduced and analyzed in Section \ref{subsec:specana}. 

\subsection{WISE MIR Photometry} \label{subsec:wise}
AT\,2022dbl has been continuously observed by the Wide-field Infrared Survey Explorer \citep[WISE;][]{Wright2010}, and the successive Near Earth Object Wide-field Infrared Survey Explorer \citep[NEOWISE;][]{Mainzer2011,Mainzer2014}, at W1 (3.4 $\mu$m) and W2 (4.6 $\mu$m) bands every half year. 

To check the potential MIR dust echo~\citep{Jiang2016,vV2016}, we query and download the W1- and W2-band photometric data from the AllWISE Multiepoch Photometry Table and the NEOWISE-R Single Exposure (L1b) Source Table. We filter out the bad data points that have 
\texttt{NaN} magnitudes and errors; or get affected by 
a nearby image artifact (\texttt{cc\_flags} $\neq$ 0), the scattered moon light (\texttt{moon\_masked} $\neq$ 0), or a nearby detection (\texttt{nb} $>$ 1). The remaining data points are grouped into approximately half-year bins to enhance the signal-to-noise ratio.
No variability has been detected in the four epochs since the rise of the previous flare (MJD $\sim$ 59706$-$60279). The averaged W1$-$W2 Vega magnitude for the host galaxy is 0.007 $\pm$ 0.006. This results is consistent with \citet{Jiang2021}, which found that most optical TDEs show very weak IR echoes likely due to a very low dust covering factor, while the W1$-$W2 color is against the AGN selection criterion: W1$-$W2 $\geqslant0.8$ \citep{Stern2012}.

\subsection{Radio Observations} \label{subsec:radio}
According to \citet{Sfaradi2022}\footnote{\url{https://www.wis-tns.org/astronotes/astronote/2022-57}}, on February 26, 2022 (around the peak of the previous flare), a 2-hour VLA observation revealed a single faint point source with flux density of 32 $\pm$ 7 $\mu$Jy in the Ku-band ($\nu\sim$ 15 GHz). The distance is $\sim$0.4 arcsec from the reported position of AT\,2022dbl, which is consistent with the position of the center of the host galaxy. However, the data are not publicly available.

Furthermore, the position of AT\,2022dbl has also been observed by the Very Large Array Sky Survey \citep[VLASS,][]{Lacy2020} for three times. Two of the observations were performed before the flare: Epoch 1.1 on November 20, 2017 and Epoch 2.1 on August 1, 2020. The other is Epoch 3.1 on February 4, 2023, which was taken $\sim$1 year after the peak of the previous flare. We retrieve tables and cutouts from the VLASS quick look catalog from CIRADA\footnote{\url{https://cirada.ca/vlasscatalogueql0}}, and confirm that no source has been detected within a radius of 1' in all three epochs. Hence, we shall not discuss the radio properties in this work.

\section{Fitting plots for optical spectra}

\begin{figure*}[htb!]
    \centering
    \figurenum{B1}
    \epsscale{0.78}
    \plotone{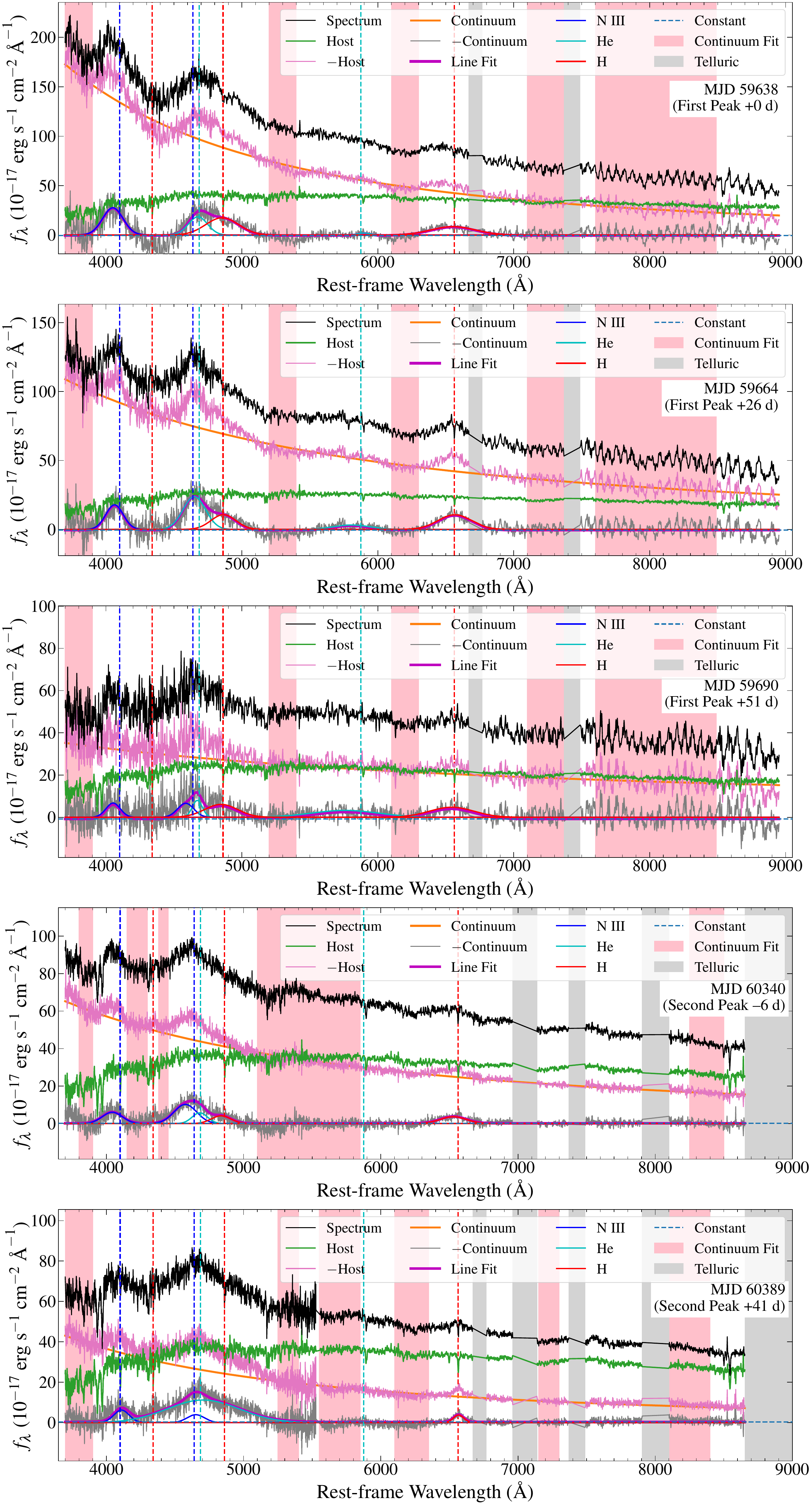}
    \caption{The fitting plots for optical spectra of AT\,2022dbl.}
    \label{fig:specfit1}
\end{figure*}

In Figure \ref{fig:specfit1}, we plot the optical spectra individually.

\vspace{10pt}

\bibliography{sample631}{}
\bibliographystyle{aasjournal}



\end{document}